\RequirePackage{fix-cm}
\documentclass[twocolumn]{svjour3}          
\smartqed  
\usepackage{graphicx}
\journalname{Current Climate Change Reports}
\begin{document}

\title{Precipitation extremes under climate change}


\author{Paul A. O'Gorman
}


\institute{P. A. O'Gorman \at
              Department of Earth, Atmospheric, and Planetary Sciences,
              Massachusetts Institute of Technology,
              77 Massachusetts Avenue,
              Cambridge, MA 02139\\
              \email{pog@mit.edu}         
}

\date{Received: date / Accepted: date}

\maketitle

\begin{abstract}

The response of precipitation extremes to climate change is considered using results from theory, modeling, and observations, with a focus on the physical factors that control the response.  Observations and simulations with climate models show that precipitation extremes intensify in response to a warming climate. However, the sensitivity of precipitation extremes to warming remains uncertain when convection is important, and it may be higher in the tropics than the extratropics. Several physical contributions govern the response of precipitation extremes. The thermodynamic contribution is robust and well understood, but theoretical understanding of the microphysical and dynamical contributions is still being developed. Orographic precipitation extremes and snowfall extremes respond differently from other precipitation extremes and require particular attention.  Outstanding research challenges include the influence of mesoscale convective organization, the dependence on the duration considered, and the need to better constrain the sensitivity of tropical precipitation extremes to warming.

\end{abstract}

\section{Introduction}

The response of precipitation extremes (heavy precipitation events) to climate
change has been the subject of extensive study because of the potential impacts
on human society and ecosystems \cite{handmer12}.  An early study using a
4-level general circulation model found that heavy daily precipitation events
become more frequent in response to elevated atmospheric CO$_2$ concentrations
\cite{gordon92}.  Numerous model studies since then have also found an
intensification of precipitation extremes with climate warming (with
important regional variations), and this has been confirmed in the
available historical record over land, as will be discussed in detail in later
sections.

Understanding of changes in 
precipitation extremes is better than for changes in other
extremes such as tornadoes \cite{kunkel13}, but large uncertainties and
research challenges remain.
If changes in dynamics and
precipitation efficiency are negligible, precipitation extremes increase with
warming because of increases in the saturation vapor pressure of water
\cite{trenberth99,trenberth03,allen02,ogorman09b}; this will be made more
precise in section 3.  However, dynamical contributions and changes in
precipitation efficiency may also play an important role.  
Mesoscale convective organization is important for the dynamics of 
precipitation extremes in the tropics (and seasonally in the midlatitudes)
but it is not resolved in global models,
while at the same time there are relatively few observational
records of tropical precipitation extremes for estimating long-term
trends and sensitivities.  At higher
latitudes, the effect of climate change on 
snowfall extremes
and freezing rain will be different from its effect on rainfall extremes and
requires further study.  In terms of
impacts, the duration of extreme precipitation events and the response of
orographic precipitation extremes are both important and are only now receiving
substantial research attention.

This paper reviews and elaborates on some of the recent research on how climate
change affects precipitation extremes, including observed changes in the
historical record (section 2), physical theory (section 3), climate-model
projections (section 4), orographic precipitation extremes (section 5),
snowfall extremes (section 6), and the duration of precipitation extremes
(section 7).  The primary focus is on the physical factors that control the
intensity of precipitation extremes in different climates.  Open questions are
discussed throughout and in section 8.

\section{Observed changes in precipitation extremes}

\begin{figure}
  \includegraphics{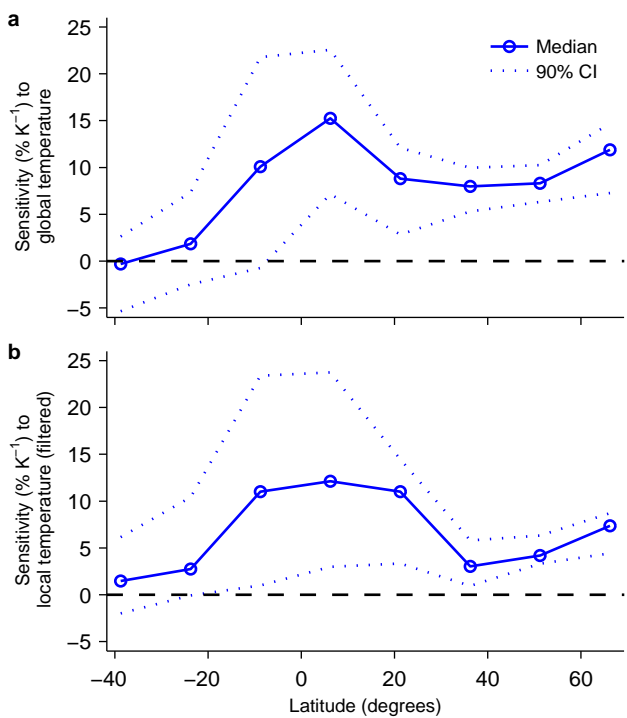}
\caption{Sensitivities of observed annual-maximum daily precipitation over land
(solid lines with circles; dotted lines show the
90\% confidence interval) 
in 15$^\circ$ latitude bands relative to (a) global-mean surface temperature
or (b) mean surface temperature over the 15$^\circ$ latitude band.
Precipitation is from HadEX2, sensitivities are calculated 
for gridboxes with at least 30 annual values,
and the median sensitivity is plotted
for each 15$^\circ$ latitude band.  Temperatures are over land and ocean from 
NOAA MLOST, and for (b) the temperature timeseries were smoothed with a 9-year running-mean filter.  
}
\label{fig_obs}
\end{figure}

Records of precipitation that are sufficient to detect long term trends in
extremes are primarily from rain gauges over land.  Over the available record,
there are regions with both increasing and decreasing trends in precipitation
extremes \cite{groisman05,alexander06},
as might be expected given large internal variability
\cite{fischer14}, but the gridboxes or stations with significant increasing
trends outnumber those with significant decreasing trends
\cite{donat13,westra13}.  Anthropogenic forcing has been
shown to have contributed to the intensification of precipitation extremes over
Northern Hemisphere land \cite{min11,zhang13}.  Assessments have also been
made of the effect of anthropogenic forcing on the probability of
specific extreme precipitation or flooding events using ensembles of
climate-model simulations \cite{pall11,otto14,herring14}.

One approach that reduces the influence of unforced variability while
still distinguishing large-scale variations 
is to analyze the sensitivity of
precipitation extremes 
averaged over all stations or grid boxes in
a latitude band \cite{westra13,asadieh15}.
Figure \ref{fig_obs}a shows an example of this type of analysis in which
annual-maximum daily precipitation rates over land
from the HadEX2 gridded dataset \cite{donat13} 
have been regressed 
over the period 1901 to 2010 against temperature anomalies
from NOAA's Merged Land-Ocean Surface Temperature Analysis 
(MLOST) \cite{smith08}.
The precipitation rates are over land only, 
but precipitation extremes do not necessarily scale with the local land mean temperature because of advection of water vapor from
over the ocean such as in atmospheric rivers
\cite{dettinger11,lavers11}, 
 and the temperatures used here are over both land and ocean.
For each gridbox with at least 30 years of data, the annual-maximum
daily precipitation rates 
are regressed against the global-mean surface temperature
anomalies using the Theil-Sen estimator, 
and the regression coefficient is divided by the mean of the
annual-maximum daily precipitation rate at the gridbox
to give a sensitivity that is expressed in units of \% K$^{-1}$.
The median of the sensitivities is then calculated for all gridboxes in
15$^\circ$ latitude bands.\footnote{ 
The circles in Fig.~1 are plotted at the midpoints of the latitude bands.
There are relatively few gridboxes for
some latitude bands, and higher-latitude bands with little
data are excluded. 
Uncertainty is estimated by bootstrapping
the years used at each gridbox (1000 bootstrap samples are generated)
and then calculating a 90\% confidence interval
for the median sensitivity in each latitude band (or averaged over several
latitude bands).
}
The resulting sensitivity is positive for most latitude bands,
the 90\% confidence interval is above zero
for all latitude bands in the Northern Hemisphere,
and the global sensitivity (averaging
over latitude bands with area weighting) is  8\% K$^{-1}$ with a 90\%
confidence interval of 5 to 10 \% K$^{-1}$.
These results, similar to those obtained previously \cite{westra13,asadieh15}, provide evidence for an intensification of annual-maximum daily
precipitation as the global-mean temperature has risen over the last century,
and at a rate that is roughly consistent with what might be expected from 
theory.  However, the meridional
structure of the sensitivities within the tropics is
sensitive to the details of the analysis (cf. \cite{westra13,asadieh15}).  

Extratropical precipitation extremes at a given latitude
occur when the atmosphere is warmer than average and are more closely
tied to mean temperatures somewhat further equatorward \cite{ogorman09a,ogorman09b,dettinger11}. However, they are still expected to respond primarily
to changes in mean
temperatures in the extratropics rather than the tropics, and 
recent warming has been greater in the northern extratropics than the
tropics.  The sensitivities shown in Fig.~\ref{fig_obs}a are based on
global-mean surface temperature and do not account for the variation in warming
with latitude.  Fig.~\ref{fig_obs}b shows an alternative analysis in which the 
annual-maximum daily precipitation rates at each gridbox are regressed against
the area-weighted mean temperature anomaly for the 15$^\circ$ latitude band
that contains the gridbox.  
The latitude-band temperature timeseries are 
filtered using a nine-year running mean prior to performing the regression.
This filtering reduces the influence of short-term variability 
in regional temperatures which has previously been found to give a different 
sensitivity of precipitation extremes than long-term climate change
\cite{ogorman12}.  The results in Fig.~\ref{fig_obs}b show  a higher
sensitivity of precipitation extremes in the tropics compared to the
extratropics, although the uncertainty in the tropics is large reflecting the
sparse data there.  The sensitivity for the tropics (30S to 30N) is 9\%
K$^{-1}$ (90\% confidence interval 6-14\% K$^{-1}$), while for the extratropics
it is 4\% K$^{-1}$ (90\% confidence interval 2-5\% K$^{-1}$).  The choice of
filter for the temperature time series affects the overall magnitudes of the
sensitivities but not whether sensitivities are higher in the tropics than the
extratropics.  Interestingly, higher sensitivities in the tropics are also
found when projections from global climate models are constrained by satellite
observations \cite{ogorman12} as discussed in section 4.2. 

\section{Theory}

To understand the response of precipitation extremes to warming, our starting
point is an approximation for the surface precipitation rate $P$ in an extreme
precipitation event,
\begin{equation} 
P \simeq - \epsilon \left\{ \omega(p) \, S(T,p) \right\}, 
\label{scaling} 
\end{equation}
where $\epsilon$ is a precipitation efficiency, $\omega$ is the vertical
velocity in pressure coordinates (negative for upward motion), $S(T,p) =
dq_s/dp|_{\theta_e^*}$ is the derivative of the saturation specific humidity
$q_s$ with respect to pressure $p$ taken at constant saturation equivalent
potential temperature $\theta_e^*$ (i.e., the derivative along a moist
adiabat), and $\{ \cdot \}$ is a mass-weighted vertical integral over the
troposphere \cite{ogorman09a,muller11a}.  All quantities in equation
(\ref{scaling}) are evaluated locally in the extreme event.  The net
condensation rate is approximated by $- \omega \, S$ either through
consideration of the condensation rate in a rising saturated air parcel
\cite{ogorman09b} or using a dry static energy budget in the tropics
\cite{muller11a}.  The precipitation
efficiency $\epsilon$ is defined as the ratio of surface precipitation 
to the column-integrated net condensation; it accounts for condensate and
precipitation storage or transport from the column.  Note that $\epsilon$ is
not a conventional precipitation efficiency because it is defined in terms of
net condensation (condensation minus evaporation) rather than condensation. 

According to equation (\ref{scaling}), changes in the precipitation rate in
extreme events under climate change have a dynamical contribution from changes
in $\omega$, a thermodynamic contribution from changes in $S$ (this is termed
thermodynamic since $S$ only depends on temperature and pressure), and a
microphysical component from changes in the precipitation efficiency
$\epsilon$.  Relative humidity does not explicitly appear in equation (\ref{scaling}), but it can affect precipitation extremes through the dynamics and by helping to set the duration of precipitation events.
The fractional increase in $S$ with warming is influenced by changes in the
moist adiabatic lapse rate \cite{betts87,ogorman09b} and varies strongly
depending on temperature and therefore altitude in the atmosphere.  However,
for a moist-adiabatic stratification and convergence confined to
near the surface, the thermodynamic contribution can be shown to scale in a
similar way to near surface specific humidities
\cite{ogorman09a,romps11,muller11a}.  This scaling is often referred to as
Clausius-Clapeyron scaling and gives a sensitivity of 6-7\% K$^{-1}$ for
typical surface temperatures.  More generally, the thermodynamic contribution
depends on the weighting of $S$ by the vertical velocity profile in the
vertical integral in equation (\ref{scaling}), and a range of higher and lower
rates of change from the thermodynamic contribution have been found in different
simulations \cite{ogorman09a,muller11a,shi15}.
It is sometimes stated that the dynamical contribution must be positive for a
warming climate because of increases in latent heating, but this is not
necessarily the case because other factors such as increases in dry static stability or reductions in meridional temperature gradients can counteract the increases in latent heating. Instead, the dynamical contribution is discussed here separately for different dynamical regimes. For example, increases in convective updraft velocities with warming are discussed in the next paragraph, and changes in large-scale vertical velocities in the extratropics are discussed in section 4.1 using the omega equation.  

The simplest configuration for which the contributions to changes in
 precipitation extremes have been analyzed is
radiative-convective equilibrium (RCE) in a doubly periodic domain
\cite{romps11,muller11a,muller13,singh14}.  
There are no large-scale dynamics
in RCE, and cloud-system resolving models (CRMs) are used to resolve the
convective-scale dynamics.  Both the convective available potential energy
(CAPE) and the updraft velocities in the middle and upper troposphere increase
with warming in RCE \cite{romps11,muller11a}; as the atmosphere warms the
thermal stratification remains close to neutral to a strongly entraining plume,
and this implies increases in CAPE (calculated based on a non-entraining
parcel) and increases in updraft velocities for more weakly entraining plumes
\cite{singh13,singh15}.  But the increases in updraft velocities in the upper
troposphere do not strongly affect the precipitation extremes because the
factor of $S(p)$ in equation (\ref{scaling}) gives more weight to the vertical
velocities in the lower troposphere in determining the intensity of
precipitation extremes.  For surface temperatures near those of the present-day
tropics, the precipitation extremes increase at close to the thermodynamic
rate, and this is close to Clausius-Clapeyron scaling with the surface specific
humidity, with relatively small contributions from changes in vertical
velocities and precipitation efficiency \cite{romps11,muller11a,singh14}. The
same behavior is found when convection is organized in a squall line
\cite{muller13}.  However, for temperatures below 295K, the precipitation
efficiency can change substantially with warming and the scaling of
precipitation extremes then depends on the accumulation period considered
\cite{singh14}, as discussed in section 7.

\section{Climate-model projections}
\label{section_models}

\begin{figure}
  \includegraphics{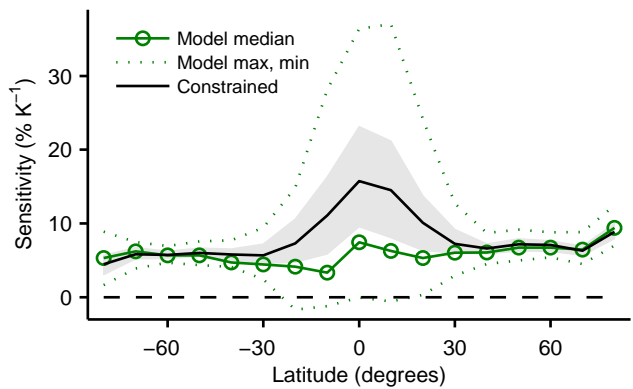}
\caption{Sensitivity of the 99.9th percentile of daily
precipitation to global-mean surface temperature for climate change
under the RCP8.5 scenario in CMIP5 global climate-model simulations.
Shown are the multimodel median (green line with circles) and
the full model range (dotted lines).
Also shown are sensitivities inferred by constraining
the model sensitivities using
observations of tropical variability (black line)
with a 90\% confidence interval 
obtained by bootstrapping as in \cite{ogorman12} (gray shading).}
\label{fig_cmip5}
\end{figure}

Climate models provide global coverage for precipitation extremes
\cite{sun07,kharin13} and more detailed coverage on regional scales
\cite{diffenbaugh05,kawazoe13,kendon14,ban14}. They may be applied to 
different emissions scenarios or individual radiative forcings 
\cite{kharin07,chen11,kao11}, 
and they allow relatively straightforward
investigations into the role of dynamics and other factors that contribute
to precipitation intensity \cite{emori05,pall07,ogorman09a,sugiyama10}.  
Important limitations in the ability of
current models to simulate precipitation extremes
have also been recognized and are related in part to the use of parameterized
convection \cite{wilcox07,kharin07,ogorman12,wehner13,kooperman14}.

Global models precipitate too frequently with too low a mean
precipitation intensity \cite{dai06,stephens10}, but this does not necessarily
mean that they underestimate the intensity of precipitation extremes.  For
example, in an analysis of 30-year return values of daily precipitation over
the conterminous United States, most global climate models were found to
overestimate or roughly agree with observations that were conservatively
interpolated to the model resolution for comparison \cite{chen08}. 
(Appropriate interpolation of precipitation is important because of
mismatches in time and space scales between models and observations.)
One exception was the Community Climate Model System 3 which underestimated the
30-year return values \cite{chen08}, and increased horizontal resolution
\cite{wehner10} or use of superparameterization \cite{li12} have been shown to
improve the representation of the intensity distribution of precipitation in
the Community Atmosphere Model versions 2 and 3.  The model bias of
too-frequent precipitation mentioned above will affect
percentiles calculated over only wet days rather than all days \cite{ban15},
even if the extreme events are properly simulated,
which suggests that calculating extremes using all days (or all hours) is
preferable for comparison of precipitation extremes between models and
observations. 

Projections of 21st century climate change with global climate models
show a general increase 
in the intensity of precipitation extremes except in some
regions in the subtropics \cite{kharin07,kharin13}.
To illustrate basic features of the response, 
Fig.~\ref{fig_cmip5} shows the sensitivity of the 99.9th percentile
of daily precipitation to warming as a function of latitude
in simulations with 15 global climate models from the 
Coupled Model Intercomparison Project phase 5 (CMIP5).
Sensitivities for climate change (\% K$^{-1}$)
are calculated as the change in the 99.9th percentile between
the final two decades
of the 20th century in the historical simulations and the final
two decades of the 21st century in the warmer RCP8.5 simulations, 
normalized by the value in the
historical simulations and the change in global-mean surface
air temperature.\footnote{The model names and the exact time periods
used are given in \cite{ogorman12}. 
The daily precipitation rates are
first conservatively interpolated \cite{chen08} to an equal-area grid with
constant spacing in longitude of 3 degrees.
Following \cite{ogorman09a}, the precipitation extremes in a given climate are 
calculated by aggregating daily precipitation rates (over both
land and ocean and including dry days) at a given latitude and then calculating
the 99.9th percentile. Calculating the change in precipitation extremes at each
gridbox and then taking the zonal average has been found to give similar
results \cite{pall07}.  
The sensitivities are averaged over 10$^\circ$ latitude bands for 
presentation in Fig.~\ref{fig_cmip5}.
}
Note that the sensitivities from observations
in Fig.~1 and from simulations in Fig.~2 should not be compared in detail
because of the different time periods, geographic coverage, and measure of
extreme precipitation used.  
We next discuss the simulated response of extratropical precipitation extremes,
followed by tropical precipitation extremes and the use of observed variability to better constrain the intermodel spread.  

\subsection{Extratropical precipitation extremes}

The multimodel-median sensitivity is shown by the green line with circles in Fig.~\ref{fig_cmip5}, and the multimodel-median of the sensitivity averaged over
the extratropics is 6\% K$^{-1}$.
A slightly lower extratropical sensitivity of 5\% K$^{-1}$ is obtained if it is
normalized by the change in extratropical-mean surface temperature rather
than global-mean surface temperature.
The intermodel spread in the response (dotted lines in Fig.~\ref{fig_cmip5}) 
is small in the extratropics, 
consistent with the fact that most precipitation extremes there
are associated with cyclones and fronts \cite{pfahl12,catto13} that may be 
expected to be reasonably well simulated.
However, global models with conventional parameterizations are unable
to simulate
precipitation extremes from mesoscale convective systems over
midlatitudes in summer \cite{kooperman14}, and so the results from these
models are not reliable for regions and times of year in which these
systems are important.

Equation (\ref{scaling}) with the precipitation efficiency $\epsilon$ held
fixed reproduces the fractional changes in precipitation extremes in CMIP3
simulations \cite{ogorman09a,sugiyama10}.  The thermodynamic contribution in
these simulations is close to what would be expected from scaling of
precipitation extremes with surface specific humidity, and this implies a lower
rate of increase than scaling with column water vapor \cite{ogorman10a}.  In
the extratropics, the simulated rate of increase of precipitation extremes is
close to the thermodynamic contribution at all latitudes, and there is little
dynamical contribution from changes in vertical velocities
\cite{emori05,ogorman09a}.  A stronger dynamical contribution
resembling a poleward shift has been found in idealized
aquaplanet simulations \cite{ogorman09b,lu14}.

Why is there no general strengthening or weakening of large-scale
vertical velocities associated with simulated
extratropical precipitation extremes 
despite changes in latent heating and dry static stability?
As a starting point,
consider the quasigeostrophic omega equation written as
\begin{equation} 
\nabla^2 \left(\sigma \omega + \frac{\kappa J}{p}\right)  + f_0^2 \frac{\partial^2}{\partial p^2} \omega = \mathrm{RHS},
\label{omega_equation}
\end{equation}
where $\omega$ is the vertical velocity in pressure coordinates,  
$\sigma$ is the dry static stability parameter, $J$ is the diabatic
heating rate, $\kappa$ is the ratio of the gas constant
to the specific heat capacity at constant pressure, 
$p$ is pressure, $f_0$ is a
reference value of the Coriolis parameter, and the right-hand-side
($\mathrm{RHS}$) includes vorticity and temperature advection terms but 
not the static stability or diabatic heating rate \cite{holton04}.
This equation is the simplest 
equation for the vertical velocity that accounts for dynamical balance, and
it is used here 
to gain some insight into the controls on large-scale
vertical velocities in the extratropics,
although it is not expected to
be quantitatively accurate.
In a strong non-convective event with
saturated ascent, $J$ will be dominated
by latent heating and $\sigma + \kappa J/ (p \omega)$ is a measure of
the moist static stability.
This moist static stability will be small if the stratification
is close to moist adiabatic, as was the case,
for example, in the extreme precipitation event in the Colorado Front
Range in September 2013 \cite{gochis15}.
For a region of upward motion that is sufficiently broad in 
the horizontal with small moist static stability,
the omega equation (\ref{omega_equation})
reduces to 
$f_0^2 \frac{\partial^2}{\partial p^2} \omega \simeq \mathrm{RHS}$,
and the effect of climate change on
the vertical velocity $\omega$ 
does not depend on 
changes in static stability
or latent heating. The vertical velocity still depends
on RHS,
but changes in this would be expected to be relatively small
given modest changes in eddy kinetic energy 
\cite{ogorman10} and eddy length \cite{kidston10}.

The omega equation (\ref{omega_equation}) gives, therefore,
some insight as to why the vertical velocities
associated with large-scale extratropical precipitation extremes
might not change greatly under climate change. 
The term proportional to
$f_0^2$ on the left hand side of equation (\ref{omega_equation})
arises from planetary rotation 
and it makes the large-scale vertical velocity
much less sensitive to deviations from a moist adiabatic stratification when
compared to small-scale convective updrafts (see sections 3 and 7).  
We next turn to the tropics where 
the dynamical influence of planetary rotation is weaker and where convection
is always a key factor for precipitation extremes.

\subsection{Tropical precipitation extremes}
\label{tropical_section}

As compared to the extratropics, the intermodel range in the sensitivity of precipitation extremes
is much larger in the tropics (Fig.~\ref{fig_cmip5}), 
with close to zero sensitivity in some
models and greater than 30\%K$^{-1}$ in others.  Additional
reasons to doubt the response of tropical precipitation extremes in these
global climate models include the large differences
between tropical precipitation extremes in 20th-century 
simulations in different models \cite{kharin07}, 
the inability of the models to represent 
mesoscale convective organization \cite{rossow13}
or to simulate the
interannual variability in tropical precipitation extremes when compared to
observations \cite{allan08,allan10}, 
and the disproportionate increases in precipitation extremes
compared to other parts of the precipitation distribution that is found in some
models -- an ``extreme mode'' in the tropical response to climate change that
relates to gridpoint storms \cite{pendergrass14a,pendergrass14b}.

Observations can be used to better constrain the large uncertainty in the
response of tropical precipitation extremes to warming.
The sensitivity of tropical precipitation extremes for
climate change in different climate models is correlated with their sensitivity
for shorter term variability within a climate (variability that is
primarily related to El Ni\~{n}o-Southern Oscillation) \cite{ogorman12}.
For example, models with a relatively high sensitivity of
tropical precipitation extremes for climate change also have a relatively high
sensitivity of tropical precipitation extremes 
for variability in historical simulations, although the sensitivities for climate change and variability are generally different in value.  
The robust relationship between the sensitivities across models has
been used together with observed variability to constrain the
sensitivity of tropical precipitation extremes to climate change
\cite{ogorman12}. The black line in Fig.~\ref{fig_cmip5} shows 
a similar observationally-constrained estimate of the
sensitivity of the 99.9th percentile of daily precipitation 
for climate change, but
instead of considering the sensitivity for climate change
aggregated over the whole tropics as in \cite{ogorman12}, 
the analysis is applied separately to the
sensitivity for climate change in 10$^\circ$ latitude bands in both the tropics
and extratropics.\footnote{
The observationally-constrained estimate 
is obtained by regressing the sensitivity for
climate change against the sensitivity for variability across the models, and
then using this regression relationship together with
the observed sensitivity for variability to
estimate the sensitivity for climate change.
The sensitivity for variability (\% K$^{-1}$) 
is calculated in both models and observations
based on the 99.9th percentile of daily precipitation
rates aggregated over the tropical oceans and the mean surface temperature over
the tropical oceans (30S to 30N) 
as described in detail in \cite{ogorman12}.
The sensitivities for climate change are calculated in 10$^\circ$ latitude
bands relative to the change in global-mean surface temperature,
as for the other sensitivities shown in
Fig.~\ref{fig_cmip5}. 
Differences in convective parameterizations
are less important in the extratropics, and
the correlation coefficient across models 
between the sensitivities for climate change and 
variability becomes smaller for climate change at higher
latitudes, reaching a value of 0.5 at 50S and 40N 
as compared to a maximum of 0.86 at 20N.
SSM/I data from Remote Sensing Systems \cite{hilburn08}
are used for the observed precipitation rates and 
NOAA MLOST data \cite{smith08}
for the observed temperatures.  
} 
This observationally-constrained estimate 
is similar to the multimodel median in the extratropics but higher than the
multimodel median in the tropics. It
peaks near the equator and is higher for the tropics
(11 \% K$^{-1}$, 90\% confidence interval 7-15 \% K$^{-1}$) than the extratropics (6 \% K$^{-1}$, 90\% confidence interval 6-7\% K$^{-1}$).  
Interestingly, a higher sensitivity in the tropics compared to
the extratropics was also found using
historical rain-gauge data (section 2). 
For the tropics, there still remains
considerable uncertainty in both the estimate from rain-gauge data and the
observationally constrained estimate discussed in this section, and
better constraining the sensitivity of tropical precipitation extremes
is an important research challenge.

\section{Orographic precipitation extremes}

Idealized simulations have recently been used to study the response of
orographic precipitation extremes to climate warming \cite{siler14,shi15} (see
also \cite{kirshbaum08} for a more general discussion).  A striking result from
these studies is that there are higher fractional changes in precipitation
extremes on the climatological
leeward slope of the mountain as compared to the windward slope.
Orographic precipitation extremes must be treated
as a special case for several reasons.
The thermodynamic contribution is influenced by the 
vertical profile of the vertical
velocity (see equation 1), 
and the shape of this profile will generally be different
over a sloped lower boundary than over a flat lower boundary
\cite{siler14}.  Downstream transport of
precipitation means that the local
precipitation efficiency can vary
strongly over the mountain, and the condensation that leads to leeward
precipitation may occur relatively high
in the atmosphere where sensitivities to
temperature change are greater \cite{siler14}.
In addition, changes in vertical velocities are governed by mountain wave
dynamics and have been found to be different for extreme precipitation events
on the western and eastern slopes of an idealized midlatitude mountain 
\cite{shi15}.

A weakening of orographic rain shadows related to changes in precipitation
extremes has previously
been noted in simulations of climate warming over North America
\cite{diffenbaugh05,singh_d13}. 
Further study is needed to assess the
role played by the physical factors discussed above in determining changes in
orographic precipitation extremes in comprehensive simulations and observations. 

\section{Snowfall extremes}

Changes in snowfall extremes have received relatively little research
attention, party because of the difficulties in producing long-term records of
snowfall. Observational studies of daily snowfall extremes have been regional
in nature and have found large interdecadal variability with, for
example, no long term trend for Canada \cite{zhang01} but more frequent
extremes snowstorms in recent decades in the eastern two-thirds of the United
States \cite{kunkel13}.  Studies using different metrics have reached different
conclusions as to whether there are more heavy snowfall events in anomalously
warm or cold years or seasons in the United States
\cite{changnon_temporal_06,kunkel13}

Physically, snowfall extremes are expected to be affected by climate warming
through both increases in saturation vapor pressures and changes in the
frequency of occurrence of temperatures below the rain-snow transition
temperature.  A simple asymptotic theory of snowfall extremes has been
developed based on the temperature dependencies of precipitation extremes and
the rain-snow transition \cite{ogorman14}.  According to the simple theory,
snowfall extremes tend to occur near an optimal temperature of roughly
-2$^\circ$C when snowfall is measured by liquid water equivalent.  The optimal
temperature arises because saturation vapor pressures increase with temperature
whereas the fraction of precipitation that falls as snow reduces sharply at
surface temperatures near freezing. When snowfall is measured by depth
of snow, the optimal temperature is lower 
(roughly -4$^\circ$C) because the variation of snow
density with temperature must also be taken into account.  For an infinitesimal
climate warming, the intensity of snowfall extremes decreases for
climatological-mean temperatures above the optimal temperature and increases
for climatological-mean temperatures below it.  Furthermore, fractional changes
in high percentiles of snowfall are smaller the higher the percentile
considered (unlike for rainfall extremes), such that fractional changes in the
intensity of the most extreme events tend to be relatively small.
There may still be large
fractional decreases in snowfall extremes with warming in regions with
climatologically mild temperatures, and changes in the frequency of exceeding a
high threshold of snowfall may still be substantial.  

Snowfall extremes in simulations with global climate models from CMIP5 behave
similarly to the simple theory for sufficiently extreme statistics
\cite{ogorman14}, although the climatological temperature below which snowfall
extremes intensify is lower than the simple theory predicts.  The response of
snowfall extremes is similar in the subset of models that most realistically
simulate Arctic sea ice \cite{ogorman14}, the decline of which has
been hypothesized to affect midlatitude weather extremes 
\cite{cohen14}.  

Regional climate-model simulations 
exhibit large fractional decreases in maximum winter daily snowfall over much
of western Europe, but little change or increases in
other parts of Europe 
that are climatologically colder \cite{devries14}. As in the simple
theory and in global climate-model simulations, there is a strong link in
regional simulations \cite{devries14} and downscaled global simulations
\cite{lute15} between the changes in snowfall extremes and the local
climatological temperature in the control climate.

\section{Duration of precipitation extremes}

\begin{figure}
  \includegraphics{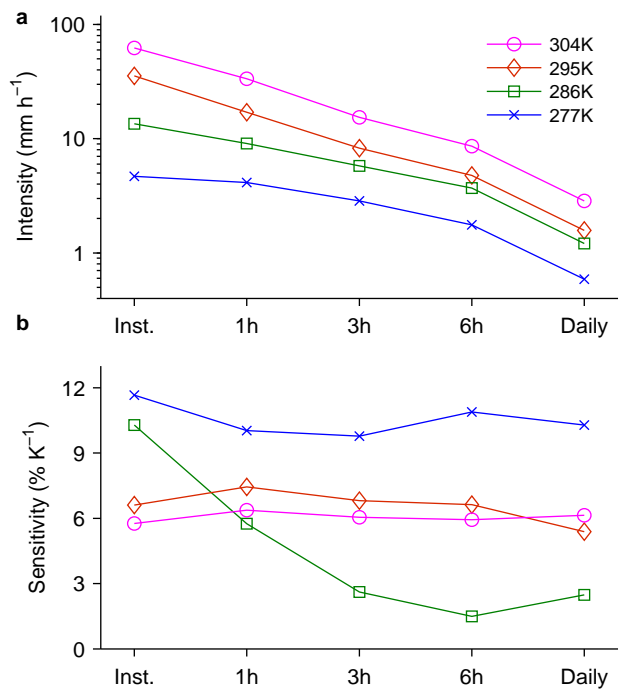}
\caption{(a) The 99.99th percentile of precipitation for different durations
(instantaneous, 1h, 3h, 6h, and daily) in simulations
of radiative-convective equilibrium with a cloud-system resolving
model at selected mean surface-air temperatures as given
in the legend. (b) The sensitivity of the
99.99th percentile of precipitation to mean surface air temperature changes
for the same temperatures shown in (a).
The natural logarithm of the 99.99th percentile of precipitation
as a function of mean surface-air temperature from 10 simulations 
is linearly interpolated to a uniform grid in temperature
and sensitivities (\% K$^{-1}$) are calculated as the change for a 3K warming.}
\label{fig_crm}
\end{figure}

The impact of changes in precipitation extremes depends
on the duration of precipitation considered (i.e., the accumulation period).
In a recent climate-model study,
intensity-duration-frequency curves were calculated for accumulation periods
from 6 hours to 10 days, and the curves were found to shift upwards in
intensity on a logarithmic scale in a relatively simple way as the climate warms
\cite{kao11}. However, it is not clear that global climate models can be relied
on for subdaily extremes because of the potential importance of convective
processes. Indeed, for regional simulations of midlatitudes in summer, changing
from a model with convective parameterization to a CRM has been found to lead
to a marked improvement in the intensity distribution of hourly
precipitation \cite{ban14} and to significantly alter the simulated response
of hourly precipitation extremes to climate change \cite{kendon14}.  

Long-term observational records of subdaily precipitation are relatively
sparse, which makes it difficult to give a general assessment of trends in
subdaily extremes \cite{westra14}.  Many recent observational studies have
instead focussed on the relationship between short-duration precipitation rates
and the local surface temperature in variability within the current climate. In
the first of these studies, a high resolution record from the Netherlands was
found to give a sensitivity of $7\% $K$^{-1}$ for daily precipitation as
compared to $14\% $K$^{-1}$ for hourly precipitation over a range of
temperatures \cite{lenderink08}. Similar behavior was found in some but not all
subsequent studies in different regions
\cite{jones10,lenderink11,utsumi11,mishra12,berg13}; see  \cite{westra14} for
an in-depth discussion.  Factors such as relative humidity
\cite{jones10,lenderink11}, large-scale dynamics and temperature gradients
\cite{mishra12}, and transitions from stratiform to convective precipitation
\cite{haerter09,berg13} are thought to be important for the scaling of
precipitation extremes with temperature in the current climate, and some of
these factors may have a different effect on hourly and daily precipitation.
While the
sensitivity of precipitation extremes for long term climate change need not be
the same as for variability in a given climate \cite{ogorman12,ban15},
understanding the sensitivity of subdaily precipitation extremes in the
present-day climate is an important starting point.

Idealized CRM studies suggest that changes in both dynamics and precipitation
efficiency could contribute to the scaling of subdaily convective 
precipitation extremes with temperature.  
Convective precipitation extremes have been found to
increase with warming considerably faster than implied by
 Clausius-Clapeyron scaling 
in some cases when a temperature increase is imposed that is
constant in the vertical \cite{singleton13,attema14}.  This is not surprising
because a vertically-uniform temperature increase makes a moist
atmosphere less statically stable and leads to faster updrafts
\cite{loriaux13}, but it does demonstrate that changes in the static
stability associated with subdaily extreme precipitation events are worthy of
further study. In a related result,
temperature changes in climate-change simulations
were found to be close to constant in the
vertical for high-CAPE
composites in the midlatitudes \cite{loriaux13,attema14}.  

As discussed in section 3, updrafts do become somewhat faster with warming when
lapse rates are allowed to equilibrate (rather than being imposed) in
simulations of RCE, although the dynamical contribution to changes in
precipitation extremes is still relatively small \cite{romps11,muller11a,singh14}.
Nonetheless, large deviations from Clausius-Clapeyron scaling have been found
in a study of RCE because of changes in precipitation efficiency at mean surface
temperatures below 295K \cite{singh14}. The 99.99th percentile of precipitation
from this study is shown in Fig.~\ref{fig_crm}a for durations from
instantaneous to daily.\footnote{Percentiles are calculated including zero
precipitation amounts, and the simulations have 500m horizontal
grid spacing and use the cloud microphysics scheme referred to as
``Lin-hail'' in \cite{singh14}.} 
Warming shifts the percentile
curves upwards in intensity in Fig.~\ref{fig_crm}a, but the rate at which
they shift upwards varies with duration and temperature.
As shown in Fig.~\ref{fig_crm}b, the
precipitation extremes follow Clausius-Clapeyron scaling at roughly 6-7\% K$^{-1}$ for temperatures above 295K.
However, for temperatures below 295K the sensitivity 
varies widely depending on temperature and accumulation period in a manner
that is not fully understood. Instantaneous precipitation extremes
increase at close to double the Clausius-Clapeyron rate for temperatures below
295K, and this has been shown to be due to
increases in precipitation efficiency with warming, related in part to
increases in hydrometeor fall speed as more of the precipitation
in the column changes from solid to liquid \cite{singh14}.
Such changes in precipitation efficiency
might be expected to occur for variability within a climate as well as for
longer term climate change, but in the simulations
they depend strongly on the choice of cloud
microphysics scheme, and it remains to be seen if they are 
relevant for observed precipitation extremes.  

\section{Conclusions and open questions}

As demonstrated in several observational studies, there has been an overall
intensification of daily precipitation extremes as a result of global warming,
although the available data has limited geographic coverage, and there are
large regional variations in the observed trends.  Much of the characterization
of projected changes in precipitation extremes comes from climate models that
use parameterized moist convection, but these are not expected to be reliable
for precipitation extremes that are primarily convective in nature 
(for example, in the tropics or for
certain events in summer in the extratropics).  As a result, simulations that
use cloud-system resolving models or superparameterizations
are becoming increasingly important to research in this area.
Even when convective dynamics are resolved, 
precipitation extremes
at short durations have been found to be sensitive to the parameterization of
cloud and precipitation microphysics \cite{singh14},
and progress in observations and physical understanding remains equally important.

Contributions from changes in thermodynamics, dynamics, and precipitation
efficiency have all been found to be important for changes in precipitation
extremes in at least some situations in modeling studies.  The thermodynamic
contribution is the easiest to understand and always gives an intensification
with warming. 
There is some basic understanding of dynamical
contributions at the large-scale from the omega equation 
(section 4.1) and also at the convective scale in the case of RCE (sections
3 and 7), but only a few studies have focussed on the role of
mesoscale convective organization in precipitation extremes 
\cite{rossow13,singleton13,muller13}.  

Precipitation extremes associated with particular dynamical regimes or
particular precipitation types may respond differently to climate warming and
are deserving of special attention. As discussed in section 5, recent idealized
studies of orographic precipitation extremes have found that 
fractional increases are larger on the climatological
leeward side than on the windward side, and
further work is needed to relate this to more realistic modeling studies and
observations.  Similarly, snowfall extremes behave quite differently from
rainfall extremes because they tend to occur near an optimal temperature
that is unaffected by climate warming.
Further work is needed to understand the specific responses of lake-effect and
high-elevation snowfall extremes, as well as changes in the frequency of hail
and ice storms \cite{changnon00,changnon03}.

Characterizing the dependence of changes in precipitation extremes on duration
is of importance for impacts, and this is particularly challenging for subdaily
durations.  Much research has focussed on precipitation accumulated over fixed
time periods as discussed in section 7.  An alternative approach is to consider
properties of contiguous precipitation events that are defined based on when
non-zero precipitation begins and ends \cite{peters10,rosa13,berg13}.
Consideration of the amount of precipitation in a given event (the event depth)
may be advantageous because observed distributions of event depths exhibit a
power law range \cite{peters10,stechmann14} and thus their response to climate
change may be relatively simple to characterize.

Daily precipitation extremes in the tropics seem to be more sensitive to
climate warming than those in the extratropics, as suggested by results from
both rain-gauge observations (section 2) and climate-model projections
constrained using satellite observations (section 4.2). One possible cause is a
more positive dynamical contribution in the tropics than the extratropics.  
Changes in extratropical eddy kinetic energy are relatively
modest and can be either positive or negative depending on season and
hemisphere \cite{ogorman10}, whereas increases in the frequency of the
most intense tropical cyclones are expected as the climate warms
\cite{knutson10}, and tropical cyclones contribute substantially to
off-equatorial precipitation extremes in the current climate \cite{lau08}.
Furthermore, increases in the activity of the Madden-Julian Oscillation and
convectively-coupled equatorial Kelvin waves have been found in simulations with
conventional and superparameterized climate models \cite{caballero10,arnold13}.
The influence of these potential dynamical changes on the aggregate statistics
of tropical precipitation extremes remains to be assessed.  

{\small
\vspace{\baselineskip}
\noindent{\bf Conflict of interest statement}\\
The author states that there is no conflict of interest. 
}

\begin{acknowledgements}

Thanks to Martin Singh, Bill Boos, Markus Donat, Lisa Alexander, and an anonymous referee for helpful comments and 
to Justin Gillis for alerting me to some early papers.
HadEX2 data were downloaded from 
www.metoffice.gov.uk/hadobs/hadex2/.
NOAA Merged Land-Ocean Surface Temperature (MLOST) (V3.5.4) data 
were provided by the 
NOAA/OAR/ESRL PSD from their website at www.esrl.noaa.gov/psd/.
SSM/I (V6) data were provided by Remote Sensing Systems
(www.remss.com) and sponsored by the NASA Earth Science MEaSUREs DISCOVER
Project.
I acknowledge the World Climate Research Programme's Working Group on Coupled Modelling, which is responsible for CMIP, and I thank the climate modeling groups for producing and making available their model output. For CMIP the U.S.  Department of Energy's Program for Climate Model Diagnosis and Intercomparison provides coordinating support and led development of software infrastructure in partnership with the Global Organization for Earth System Science Portals.
I am grateful for support from NSF grant AGS-1148594 and NASA grant NNX-11AO92G.
This paper was partly written while I was a visiting fellow at the
Climate Change Research Centre of the University of New South Wales.

\end{acknowledgements}

\section*{References}
Recent papers of particular interest (published since 2009) have been highlighted as:\\
** Of importance\\
*** Of major importance

\end{document}